\documentclass[10pt, a4paper, twocolumn, twoside]{article} % 10pt font size (11 and 12 also possible), A4 paper (letterpaper for US letter) and two column layout (remove for one column)

\usepackage[english]{babel} % English language hyphenation

\usepackage{microtype} % Better typography

\usepackage{amsmath,amsfonts,amsthm} % Math packages for equations

\usepackage[svgnames]{xcolor} % Enabling colors by their 'svgnames'

\usepackage[small, labelfont=bf, up]{caption} % Custom captions under/above tables and figures

\usepackage{booktabs} % Horizontal rules in tables

\usepackage{lastpage} % Used to determine the number of pages in the document (for "Page X of Total")

\usepackage{graphicx} % Required for adding images

\usepackage{enumitem} % Required for customising lists
\setlist{noitemsep} % Remove spacing between bullet/numbered list elements

\usepackage{sectsty} % Enables custom section titles
\allsectionsfont{\usefont{OT1}{phv}{b}{n}} % Change the font of all section commands (Helvetica)

\usepackage{geometry} % Required for adjusting page dimensions

\usepackage[T1]{fontenc} % Output font encoding for international characters
\usepackage[utf8]{inputenc} % Required for inputting international characters

\usepackage{XCharter} % Use the XCharter font

\usepackage{hyperref}

\usepackage{journals}

\usepackage{fancyhdr} % Needed to define custom headers/footers
\newcommand{\shorttitle}[1]{\fancyhead[CE]{\textsl{#1}}}
\newcommand{\shortauthors}[1]{\fancyhead[CO]{\textsl{#1}}}

\fancypagestyle{firstpage}{ % Page style for the first page with the title
	\fancyhf{}
        \lhead{\vspace*{0.0em} \textit{\footnotesize 21$^{\rm st}$ European Workshop on White
            Dwarfs \\ Castanheira, Campos, Montgomery, eds. \\[-0.2em]
          July 23--27, 2018, Austin, Texas, USA}}
}

\date{}

\newcommand{\authorstyle}[1]{{\large\usefont{OT1}{phv}{b}{n}\color{DarkRed}#1}} % Authors style (Helvetica)

\newcommand{\institution}[1]{{\footnotesize\usefont{OT1}{phv}{m}{sl}\color{Black}#1}} % Institutions style (Helvetica)
\usepackage{titling} % Allows custom title configuration

\newcommand{\HorRule}{\color{DarkGoldenrod}\rule{\linewidth}{1pt}} % Defines the gold horizontal rule around the title

\pretitle{
	\vspace{-30pt} % Move the entire title section up
	\HorRule\vspace{10pt} % Horizontal rule before the title
	\fontsize{16}{12}\usefont{OT1}{phv}{b}{n}\selectfont % Helvetica
	\color{DarkRed} % Text colour for the title and author(s)
}

\posttitle{\par\vskip 10pt} % Whitespace under the title

\preauthor{} % Anything that will appear before \author is printed

\postauthor{ % Anything that will appear after \author is printed
	\vspace{0pt} % Space before the rule
	{\par\HorRule} % Horizontal rule after the title
	\vspace{-10pt} % Space after the title section
}

\newcommand{\newabstract}[1]{
    {\section*{Abstract}
    \bfseries #1}
  }

\usepackage[authoryear]{natbib}
\title{Toward a systematic cartography of the chemical stratification inside
white dwarfs from deep asteroseismic probing of ZZ Ceti stars} % The article title

\shorttitle{Chemical stratification in ZZ Ceti stars from asteroseismology} % The short article title for page headings

\shortauthors{Charpinet, Giammichele, et al.} % The short author list for page headings

\author{
        \authorstyle{S.~Charpinet,$^1$ N.~Giammichele,$^1$ P.~Brassard,$^2$
        G.~Fontaine,$^2$ P.~Bergeron,$^2$ W.~Zong,$^3$ V.~Van~Grootel,$^4$
        and A.~S.~Baran$^5$}
	\newline\newline % Space before institutions
	$^1$\institution{IRAP, Université de Toulouse, CNRS, UPS, CNES, 14 avenue Edouard Belin, F-31400 Toulouse, France;
          [stephane.charpinet, noemi.giammichele]@irap.omp.eu}\\ % Institution 1
	$^2$\institution{Département de Physique, Université de Montréal, Montréal, QC H3C 3J7, Canada;
          [brassard, fontaine, bergeron]@astro.umontreal.ca}\\ % Institution 2
	$^3$\institution{Department of Astronomy, Beijing Normal University, Beijing 100875, P. R. China;
          weikai.zong@bnu.edu.cn}\\ % Institution 3
	$^4$\institution{Space sciences, Technologies sciences, Technologies and Astrophysics Research (STAR)
	                 Institute, Université de Liège, 19C Allée du six-août, B-4000 Liège, Belgium;
	                 valerie.vangrootel@uliege.be}\\ % Institution 3
	$^5$\institution{Uniwersytet Pedagogiczny, Obserwatorium na Suhorze, ul. Podchorżych 2, 30-084 Kraków, Polska;
	                 andysbaran@gmail.com} % Institution 3
      }

\begin{document}

\maketitle % Print the title

\thispagestyle{firstpage} % Apply the page style for the first page

%----------------------------------------------------------------------------------------
%	ABSTRACT
%----------------------------------------------------------------------------------------

\newabstract{
  DA-type white dwarfs account for 80\% of all white dwarfs and represent, for
most of them, the ultimate outcome of the typical evolution of low-to-intermediate
mass stars. Their internal chemical stratification is strongly marked by passed,
often uncertain, stellar evolution processes that occurred during the helium
(core and shell) burning phases, i.e., from the horizontal branch through AGB
and post-AGB stages. Pulsating white dwarfs, in particular the "cool" DA-type
ZZ Ceti variables, offer an outstanding opportunity to dig into these stars
by fully exploiting their asteroseismic potential. With our most recent tools
dedicated to that purpose, we show that a complete cartography of the
stratification of the main constituents of a white dwarf can be inferred,
leading in particular to strong constraints on the C/O core structure
produced by the processes mentioned above. This opens up the way toward a
systematic exploration of white-dwarf internal properties.
}

%----------------------------------------------------------------------------------------
%	ARTICLE BODY
%----------------------------------------------------------------------------------------

\section{Introduction}

Asteroseismology of white dwarf stars has long carried out the promise of revealing
the innermost structure of low-to-intermediate mass stellar remnants in their
ultimate stage of evolution. Comprehensive reviews of more than three decades
of this venture can be found in, e.g., \citet{fontaine08}, \citet{winget08},
\citet{althaus10}, and more recently through the critical discussion
provided in Section 2.1 of \citet{giam17a}. A significant breakthrough in
this field has recently been triggered by the introduction of a new approach
to obtain seismic model solutions for white dwarf pulsators that unravel their
internal chemical stratification and match better the observed period spectra
\citep{giam17a, giam17b}. This technique was successfully applied to the pulsating
DB white dwarf KIC08626021, leading to the seismic cartography of the distribution
of helium, carbon, and oxygen inside the star \citep{giam18}. Unraveling the
chemical stratification inside a white dwarf can have profound repercussions as
it bears the signatures of all the physical processes that occurred during the
evolution of stars in the helium core burning phase and beyond. Yet, one could
argue that DB white dwarfs account for only 20\% of the white dwarf population
and may have various non-typical origins (\citealt{reindl14,reindl15}; see also
Giammichele et al. in these proceedings). In contrast, the dominant DA white
dwarf population may be more representative of the outcome of main evolution
processes affecting low-to-intermediate stars in their terminal stages.
In this paper, we report on preliminary results of detailed seismic analyses
conducted on five DA-type ZZ Ceti pulsators following the same approach as
developed in \citet{giam18}.

\section{Seismic models for\\ DA pulsators}

The technique used in this study is based on a forward modeling strategy.
The adiabatic pulsation periods of white dwarf models in hydrostatic equilibrium,
whose elements of the structure such as the chemical stratification can be
parameterized with some flexibility, are calculated and quantitatively evaluated
(through a merit function) relative to the observed period spectrum of a given star.
After a global optimization procedure in the specified parameter space, the
best match seismic model(s) is (are) found. This procedure, along with the
tools developed to carry out this search, have been described at length in
\citet{giam16,giam17a,giam17b,giam18} and we refer the interested reader to
these contributions for further detail. Of relevance here, we just point out
that an updated set of DA white-dwarf models was constructed to cover the main
characteristics of the chemical composition profiles expected from evolution
calculations. These models incorporate the core parametrization introduced in
\citet{giam17a} that allows us to construct flexible oxygen profiles. They
also include the parametrized double layered helium stratification described
in \citet{giam18}, and, of course, a parametrization of the hydrogen-rich top
layer that is present in DA white dwarfs, but not in DBs.

\begin{figure}[t]
  \centerline{\includegraphics[width=1.0\columnwidth]{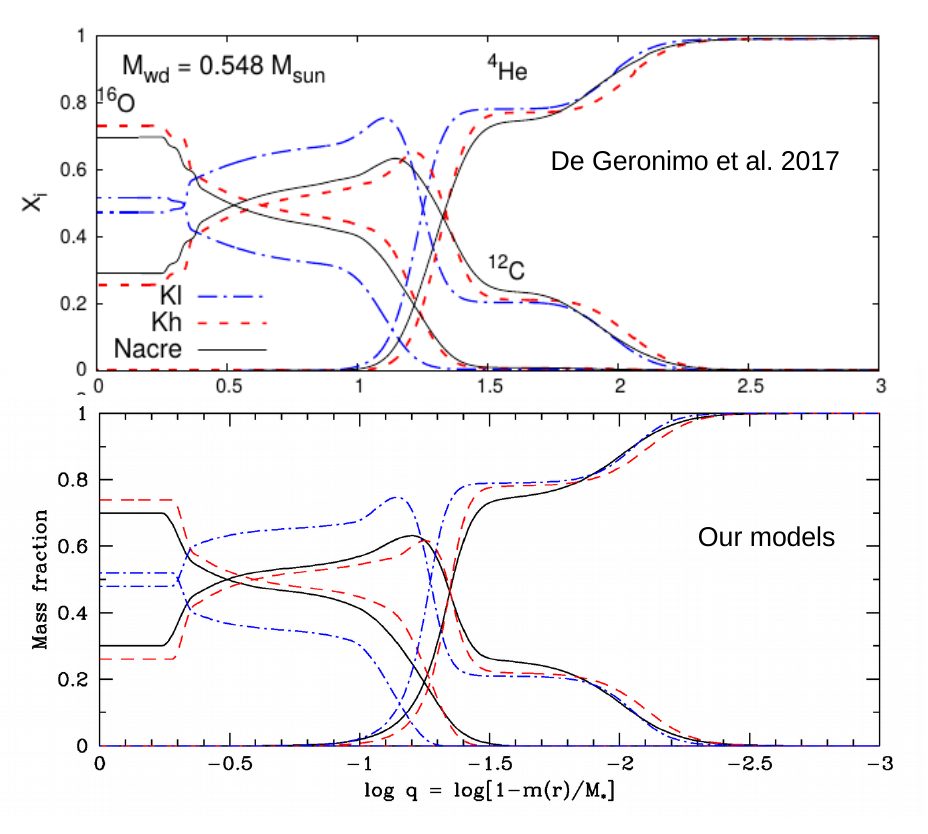}}
  \caption{Composition profiles of Helium, Carbon, and Oxygen as functions
  of the fractional mass depth in a typical DA white dwarf produced by
  assuming different nuclear reaction rates in evolution calculations.
  {\sl Top panel} is from Figure 7 of \citet{geronimo17}, while {\sl bottom panel}
  shows a similar -- eyeball fit -- reproduction of these profiles using
  our parametrized hydrostatic models developed for asteroseismology.}
  \label{fig1}
\end{figure}

\begin{figure}[t]
  \centerline{\includegraphics[width=1.0\columnwidth]{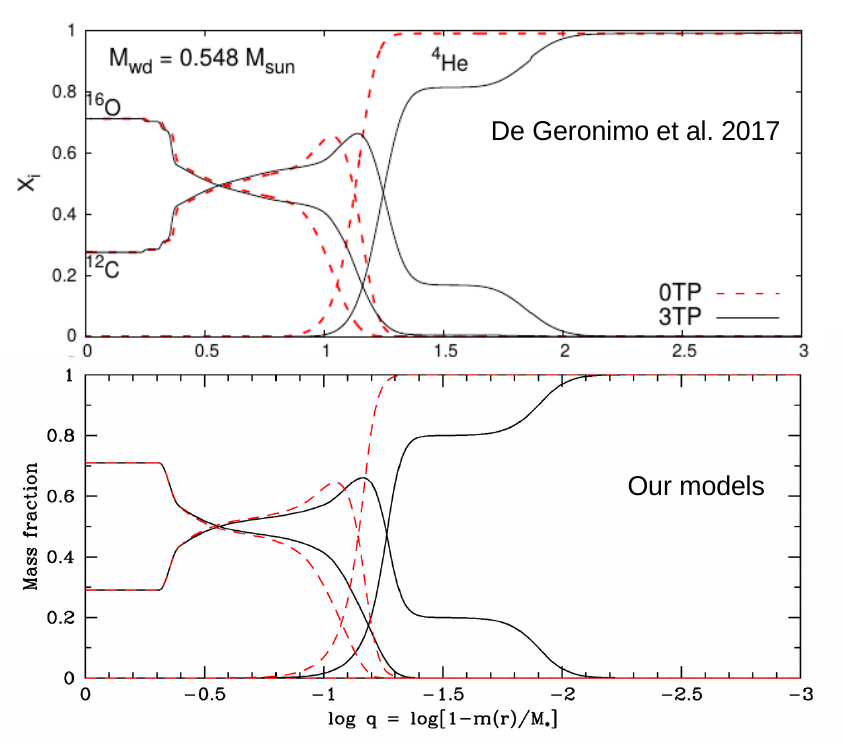}}
  \caption{Composition profiles of Helium, Carbon, and Oxygen as functions
  of the fractional mass depth in a typical DA white dwarf produced
  when no thermal pulses or 3 thermal pulses occur during the post-AGB phase.
  {\sl Top panel} is from Figure 1 of \citet{geronimo17}, while {\sl bottom panel}
  shows a similar -- eyeball fit -- reproduction of these profiles using
  our parametrized hydrostatic models developed for asteroseismology.}
  \label{fig2}
\end{figure}

\begin{figure}[t]
  \centerline{\includegraphics[width=1.0\columnwidth]{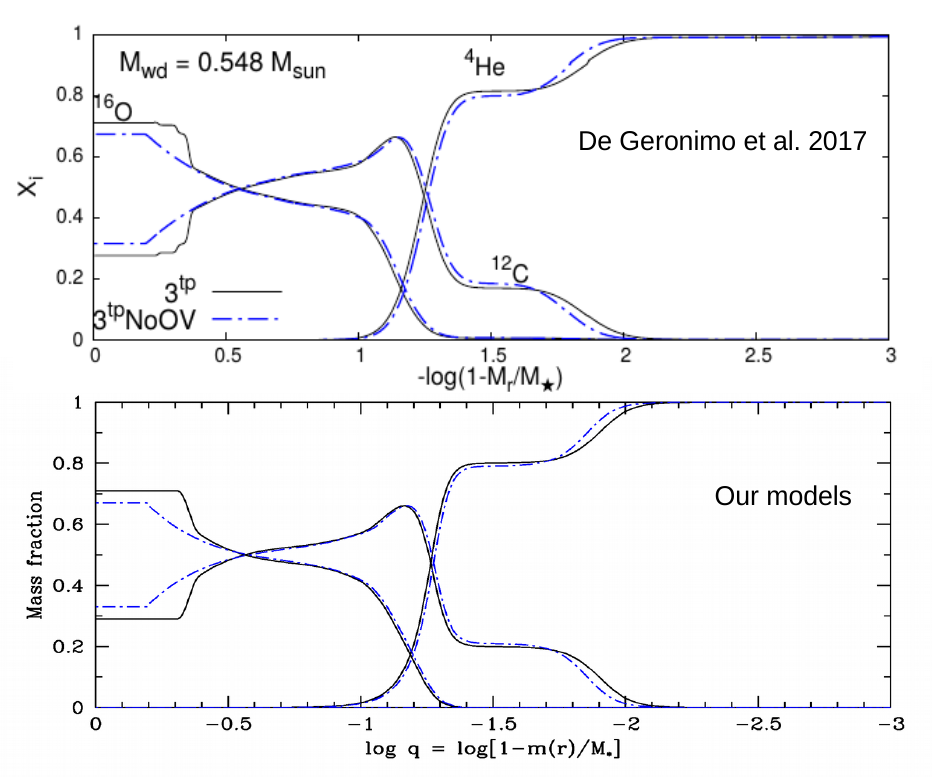}}
  \caption{Composition profiles of Helium, Carbon, and Oxygen as functions
  of the fractional mass depth in a typical DA white dwarf produced with and
  without overshoot in evolution calculations.
  {\sl Top panel} is from Figure 4 of \citet{geronimo17}, while {\sl bottom panel}
  shows a similar -- eyeball fit -- reproduction of these profiles using
  our parametrized hydrostatic models developed for asteroseismology.}
  \label{fig3}
\end{figure}

In order to illustrate the ability of these DA static models to reproduce various
chemical stratifications, including structures predicted by stellar evolution
calculations, we compare in Fig.~\ref{fig1}, \ref{fig2}, and \ref{fig3} the
profiles of C, O, and He derived from evolution models, based on \citealt{geronimo17}
calculations assuming various input physics (see figure captions),
with our static DA models adjusted to
reproduce their evolutionary counterpart. These comparisons show that our static structures can
closely mimic the predicted composition profiles, meaning that, in a context of
optimizing the stratification to find the one that best reproduces the observed
period spectrum, "typical" configurations predicted by evolution models are covered
in the search space and could potentially be found, if optimal. Of course, the
space of explored configurations is much wider as the ultimate goal of our asteroseismic
analyses is to let the oscillation periods determine what the optimal composition
profiles should be. The composition profiles derived this way can therefore potentially
differ significantly from the expected evolution profiles, but represents seismic
inversions that are essentially independent of evolution calculations.

\section{Preliminary results for\\ five ZZ Ceti stars}

We applied our optimization technique, using the new static DA white dwarf models,
to 5 ZZ Ceti stars that we estimated were promising candidates. These stars are
KIC 11911480, L19-2, SDSS J1136+0409, EPIC 220258806, and EPIC 220347759. Except
for L19-2 whose analysis relies on data from a former ground based multisite
campaign \citep{sullivan00}, all these stars have high quality photometric
lightcurves obtained from space with {\sl Kepler} (the original mission for
KIC 11911480 and K2 for the other ones).
Prior to attempting the asteroseismic fits, we reanalyzed the lightcurves
for all the {\sl Kepler} and K2 objects. In particular, new photometry extraction
from the raw K2 data was performed using a pipeline developed by one of us
(A. Baran), which in some cases provided noticeable improvements on the quality
of the extracted time series over published data. This process resulted in
isolating 5 independent periods for the seismic fit of KIC 11911480, while
L19-2, SDSS J1136+0409, EPIC 220258806, and EPIC 220347759 were analyzed based
on 5, 8, 10, and 7 independent modes, respectively. These stars being close
to the blue edge of the DAV instability strip, they all show low-degree and
low-order g-modes, and the number of modes available for the analysis proved
sufficient to obtain non-ambiguous seismic solutions in all cases (in line
with Hare \& Hound experiments discussed in \citealt{giam17b}).

\begin{table*}[t]
  \caption{Preliminary estimates of some parameters derived from asteroseismology
  for the 5 ZZ Ceti pulsators considered. When relevant, values are compared
  with available estimates from other independent methods : $T_{\rm eff}$
  (in K) and $\log g$ (in cgs) from spectroscopy, and the distance $d$ (in
  parsec) from Gaia DR2 parallax measurements.}
  \label{tab1}
  \begin{center} \begin{tabular}{llllll}
                             & KIC11911480      & L19-2            & SDSSJ1136+0409   & EPIC220258806     & EPIC220347759    \\
\hline\\
    $T_{\rm eff}$ (spectro)  & 12,026 $\pm$ 195   & 12,058 $\pm$ 184    & 12,330 $\pm$ 260  & 12,807 $\pm$ 219     & 12,692 $\pm$ 214  \\
    $T_{\rm eff}$ (astero)   & 12,300 $\pm$ 485   & 12,021 $\pm$ 717    & 12,297 $\pm$ 195  & 12,163 $\pm$ 417     & 12,478 $\pm$ 329  \\
    $\log g$ (spectro)       & 8.00   $\pm$ 0.05  & 8.11   $\pm$ 0.05   & 7.99   $\pm$ 0.06 & 8.1    $\pm$ 0.05    & 8.09   $\pm$ 0.05 \\
    $\log g$ (astero)        & 8.01   $\pm$ 0.04  & 8.13   $\pm$ 0.06   & 8.06   $\pm$ 0.04 & 8.07   $\pm$ 0.02    & 8.11   $\pm$ 0.05 \\
    $d$ (astero)             & 184.2  $\pm$ 14.6  & 21.0   $\pm$ 0.5    & 127.7  $\pm$ 7.2  & 79.6   $\pm$ 3.0     & 141.7  $\pm$ 7.0  \\
    $d$ (parallax)           & 182.9  $\pm$ 4.1   & 20.93  $\pm$ 0.01   & 130.7  $\pm$ 1.9  & 80.56  $\pm$ 0.54    & 151.8  $\pm$ 3.5  \\
\\
\hline
    \multicolumn{6}{c}{Other relevant parameters derived from asteroseismology (see text for details)}\\
\hline\\
    Mass ($M_\odot$)         &  0.63           &  0.68            &  0.63            &  0.64             &  0.67             \\
    $\log q({\rm H})$        & -3.12           & -4.39            & -5.55            & -4.23             & -4.42             \\
    $\log q({\rm He})$       & -1.42           & -1.87            & -1.75            & -1.91             & -2.40             \\
    $\log q({\rm core})$     & -0.66           & -0.80            & -0.77            & -0.37             & -0.74             \\
    O(core)                  &  0.85           &  0.82            &  0.75            &  0.88             &  0.85             \\
\\
\hline

    \end{tabular} \end{center}

\end{table*}

Detailed results from these analyses will be reported elsewhere. Here,
we provide preliminary
seismic estimates for the most relevant parameters derived for the 5 stars
considered (see Table~\ref{tab1}). In all cases, we find seismic solutions that
show a very good consistency with independent measurements of $T_{\rm eff}$
and $\log g$ from spectroscopy. The latter were derived from our most recent
grids of DA model atmospheres \citep{bergeron95,gianninas11} and by applying
the 1D/3D correction from \citet{tremblay13}. Another independent way to test
seismic solution's accuracy is now available with high precision distance
estimates provided by parallax measurements from the GAIA Data Release 2.
These distances from parallax are given in Table~\ref{tab1} and can be
compared with "asteroseismic" distances computed from the optimal seismic
model solutions. The uncovered seismic model, coupled with a corresponding
model atmosphere, provides the absolute magnitude of the star that can be
computed for any chosen filter. These absolute magnitudes can be used to estimate
interstellar reddening, extinction, and ultimately the distance modulus when
compared with measured apparent magnitudes. The values
given in Table~\ref{tab1} for the asteroseismic distance are based on the
Gaia Bp and Rp photometry and are corrected from extinction. In all cases,
we find that the distances estimated from the seismic solutions are in remarkable
agreement with those derived directly from the parallax. This result provides
a very strong test validating the seismic models obtained for these stars.

The seismic solutions uncovered also lead to a complete mapping of the chemical
stratification inside these stars. These will be disclosed in detail in
forthcoming papers (Giammichele et al. 2019, in preparation) and we restrict
this report to the
preliminary values summarized in Table~\ref{tab1}. In this table, we provide
the estimated masses for the hydrogen-rich envelope ($\log q(\rm H) = \log[1-M(H)/M_*]$),
the helium mantle, $\log q(\rm He)$, and the central homogeneous C/O core,
$\log q(\rm core)$. We also indicate the oxygen mass fraction in that central
region, $O(\rm core)$.

Our results show a rather large scatter of values for the hydrogen envelope
masses, while the derived helium layer mass fractions are in a relatively
narrow range. Remarkably, the derived central core extent
is found to be quite similar for four of these stars (out of five; see below),
ranging from $\log q(\rm core)\sim -0.66$ to $-0.80$. We emphasize that
a similar core size has also been obtained from the seismic analysis of
the DB white dwarf KIC 08626021, with $\log q(\rm core) = -0.72\pm 0.03$
\citep{giam18}.
These values lead to a mass for the central homogeneous C/O core -- mainly produced
by the former helium burning core -- that is $\sim 40\%$ larger than predicted by
standard stellar evolution calculations. This remarkable consistency
obtained for 5 white dwarfs so far strongly strengthen the conclusion that
the helium burning phase generally produces much bigger cores than currently
expected. Similarly, we find a convergence from all the white dwarfs we have
analyzed toward a larger (by $\sim 15\%$) mass fraction of oxygen produced in
the core compared to current model predictions.

Finally, we emphasize the cases of EPIC 220258806 and EPIC 220347759. These
stars turn out to be spectroscopic twins, but are clearly {\sl not} seismic
twins. Hence, while they look very similar externally, there must be important
differences in their interior structure to cause significant variations in
their oscillation period spectra. Our detailed analyses show that the
outermost layers of these stars, namely their H-rich envelope and He mantle
structure, are very similar (see again Table~\ref{tab1}). However, the
innermost parts of their core chemical stratification are significantly
different. EPIC 220347759 appears to have a rather extended central homogeneously
mixed core, comparable in size to those inferred for KIC 11911480, L19-2, and
SDSS J1136+0409. This region turns out to be much smaller in EPIC 220258806,
which therefore stands out, for this particular aspect of its structure, as
an outlier among this sample of 5 DA stars. It is not clear at this stage what
difference in the past evolution of these two stars could lead to such a
variation in their internal core structure.

\section{Conclusion}

In this study, we have applied our newly developed approach to cartography
the internal chemical stratification inside white dwarfs to 5 DA pulsators.
The optimal seismic solutions inferred, although still at a preliminary
stage, show reproducible patterns emerging from the comparison of the chemical
stratifications determined for these objects. In particular, the size and composition
of the innermost homogeneously mixed core are found to be similar for
4 stars of this sample. These are also comparable to the inner core structure
uncovered for the DB white dwarf KIC 08626021 \citep{giam18}.
This finding adds strength to the claim that the inner
part of the C/O core, produced mainly during the helium core burning phase
when a low-to-intermediate mass star evolves through the Horizontal Branch,
is much larger than presently predicted by stellar evolution calculations.
This points toward shortcomings in the treatment of mixing processes in the
core during this intermediate phase of the evolution of stars.
Yet, we found an outlier of this picture with the star EPIC 220258806,
that seems to have a reduced homogeneously mixed core compared to the other
pulsating white dwarfs analyzed thus far. Which physical process could account
for this singularity is not clear at this stage.

Overall, these results demonstrate the high potential of white dwarf asteroseismology
to shed light on still uncertain aspects of stellar structure and evolution.
With the space telescopes K2 and TESS, which have already observed or will
monitor many more pulsating white dwarfs, we expect to extend further this
type of analysis
and achieve a systematic cartography of the chemical stratification inside white
dwarfs. This journey to the center of white dwarf stars should enlighten our
understanding of these objects and have repercussions in many connected fields.

\subsection*{Acknowledgment}

We acknowledge support from the Agence Nationale de la Recherche (ANR, France)
under grant ANR-17-CE31-0018, funding the INSIDE project. This work was
granted access to the high-performance computing resources of the CALMIP
computing centre under allocation number 2018-p0205. G.F. acknowledges
the contribution of the Canada Research Chair Program, and W.Z. the
LAMOST fellowship as a young researcher, supported by the Special
Funding for Advanced Users, budgeted and administrated by the Center
for Astronomical Mega-Science, Chinese Academy of Sciences. V.V.G. is
an F.R.S.-FNRS Research Associate. The authors acknowledge the Kepler
team and everyone who has contributed to making this mission possible.
Funding for the Kepler mission is provided by NASA’s Science Mission
Directorate.

%----------------------------------------------------------------------------------------
%	BIBLIOGRAPHY
%----------------------------------------------------------------------------------------

% There are two ways to include references. The first uses bibtex and
% is recommended. For this case uncomment the following line.
%\bibliography{papers}
% Here we have assumed that the bibliography file is named
% "papers.bib"

% The second way of including references is shown below.

%----------------------------------------------------------------------------------------

\end{document}